\documentclass[11pt,twoside]{article}

\usepackage{a4wide}
\usepackage{amsfonts}
\usepackage{amssymb}
\usepackage{amsmath}
\usepackage{graphicx}

\newcommand{\ignore}[1]{}

\def\@begintheorem#1#2{\par\bgroup{\sc #1\ #2. }\it\ignorespaces}
\def\@opargbegintheorem#1#2#3{\par\bgroup{\sc #1\ #2\ (#3). } \it\ignorespaces}
\def\@endtheorem{\egroup}
\newtheorem{theorem}{Theorem}[section]
\newtheorem{corollary}[theorem]{Corollary}
\newtheorem{lemma}[theorem]{Lemma}

\newtheorem{example}[theorem]{Example}
\newtheorem{proposition}[theorem]{Proposition}
\newtheorem{definition}[theorem]{Definition}
\newcommand{\bt}[1]{\begin{theorem}\label{#1}}
\newcommand{\bc}[1]{\begin{corollary}\label{#1}}
\newcommand{\bl}[1]{\begin{lemma}\label{#1}}
\newcommand{\be}[1]{\begin{example}\label{#1}}
\newcommand{\bp}[1]{\begin{proposition}\label{#1}}
\newcommand{\ba}[1]{\begin{algorithm}\rm\label{#1}}
\newcommand{\bd}[1]{\begin{definition}\rm\label{#1}}{\normalsize }
\newcommand{\bpr}{\noindent {\em Proof. }}
\newcommand{\et}{\end{theorem}}
\newcommand{\ec}{\end{corollary}}
\newcommand{\el}{\end{lemma}}
\newcommand{\ee}{\end{example}}
\newcommand{\ep}{\end{proposition}}
\newcommand{\ed}{\end{definition}}
\newcommand{\epr}{{\ \vbox{\hrule\hbox{%
\vrule height1.3ex\hskip0.8ex\vrule}\hrule}}\\\par}

\def\R{\mathbb{R}}
\def\Z{\mathbb{Z}}
\def\conv{{\rm conv}}
\def\poly{{\rm poly}}

\begin{document}

\title{\bf The Complexity of Vector Partition}

\author{
Shmuel Onn
\thanks{\small Technion - Israel Institute of Technology.
Email: onn@technion.ac.il}
}
\date{}

\maketitle

\begin{abstract}
We consider the {\em vector partition problem}, where $n$ agents, each with a $d$-dimensional
attribute vector, are to be partitioned into $p$ parts so as to minimize cost which is a given
function on the sums of attribute vectors in each part. The problem has applications in a
variety of areas including clustering, logistics and health care. We consider the complexity
and parameterized complexity of the problem under various assumptions on the natural
parameters $p,d,a,t$ of the problem where $a$ is the maximum absolute value of any attribute
and $t$ is the number of agent types, and raise some of the many remaining open problems.


\end{abstract}

\section{Introduction}
\label{Introduction}

The {\em vector partition problem} is the following. We are given a set $[n]:=\{1,\dots,n\}$
of $n$ agents, a $d\times n$ attribute matrix $A$ whose $j$-th column $A^j$ is a
$d$-dimensional attribute vector associated with agent $j$, a positive integer $p$,
subsets $B_1,\dots,B_p\subseteq[n]$, and a function $f:\R^{d\times p}\rightarrow\R$.
A {\em $p$-partition} of $[n]$ is a $p$-tuple $\pi=(\pi_1,\dots,\pi_p)$ of sets with
$[n]=\uplus_{k=1}^p\pi_k$. A partition is {\em admissible} if $|\pi_k|\in B_k$ for all $k$,
and its cost is $f\left(\sum_{j\in\pi_1}A^j,\dots,\sum_{j\in\pi_p}A^j\right)$,
that is, the value of $f$ on the $d\times p$ matrix whose $k$-th column is the sum
of the attribute vectors of the agents in part $\pi_k$ of $\pi$.
The problem is to find a $p$-partition of minimum cost, satisfying the bounds.

\vskip.2cm\noindent{\bf Vector Partition Problem.}
Given $A\in\Z^{d\times n}$, $B_1,\dots,B_p\subseteq[n]$, and $f:\R^{d\times p}\rightarrow\R$, solve
$$\min\left\{f\left(\sum_{j\in\pi_1}A^j,\dots,\sum_{j\in\pi_p}A^j\right)
\ :\ \pi=(\pi_1,\dots,\pi_p)\ ,\ \ \uplus_{k=1}^p\pi_k=[n]
\ ,\ \ |\pi_k|\in B_k\ ,\ \ k\in[p]\right\}\ .$$

The problem is called {\em bounded-shape} if all $B_k$ are intervals, that is, $B_k=[l_k,u_k]\cap\Z$
with $0\leq l_k\leq u_k\leq n$ given lower and upper bounds for all $k$, so $\pi$ is admissible
if $l_k\leq|\pi_k|\leq u_k$ for all $k$. In particular, it is {\em single-shape} if $l_k=u_k$
for all $k$, so $\pi$ is admissible if $|\pi_k|=l_k=u_k$ for all $k$, and {\em free}
if $l_k=0$ and $u_k=n$ for all $k$, so all $p$-partitions $\pi$ are admissible.

A special case is the following {\em type partition problem}. We are now given a $t$-partition
$\tau=(\tau_1,\dots,\tau_t)$ of $[n]$ where $\tau_i$ is the set of agents of type $i$,
$B_1,\dots,B_p\subseteq[n]$, and a function $g:\R^{t\times p}\rightarrow\R$.
The cost of $\pi$ is $g\left(|\pi_1\cap\tau|,\dots,|\pi_p\cap\tau|\right)$
where $|\pi_k\cap\tau|:=(|\pi_k\cap\tau_1|,\dots,|\pi_k\cap\tau_t|)$,
that is, the value of $g$ on the $t\times p$ matrix whose $k$-th column counts the agents
of each type in part $\pi_k$ of $\pi$. To encode it as a vector partition problem
set $d:=t$, $f:=g$, and define an attribute matrix $A$ by $A^j:={\bf 1}_i$ the $i$-th
unit vector in $\R^d$ for $i=1,\dots,t$ and $j\in\tau_i$, giving
$$g\left(|\pi_1\cap\tau|,\dots,|\pi_p\cap\tau|\right)
\ =\ f\left(\sum_{j\in\pi_1}A^j,\dots,\sum_{j\in\pi_p}A^j\right)\ .$$

\vskip.2cm
Both problems have applications in a variety of areas including logistics, health care,
and clustering, see for instance \cite{COR,EBBA,HOR,OSc} and the references therein.
In particular, \cite{EBBA} describes a very recent application to group testing for COVID-19,
briefly discussed in Section \ref{Application}. They also bare relations to
constructing universal Gr\"obner bases, see \cite{BOT,OSt} and \cite[Chapter 6]{Onn}.

\vskip.2cm
In the vector partition problem we also say that two agents are of the same type if they
have the same attribute vector, and we let $t:=|\{A^j:j\in[n]\}|$ denote the number of agent types.
We also let $a:=\|A\|_\infty=\max_{i,j}|A^j_i|$ be the maximum absolute value of any attribute.

The vector partition problem is called {\em separable} if $f(x^1,\dots,x^p)=\sum_{k=1}^p f_k(x^k)$
for some given functions $f_k:\R^d\rightarrow\R$, and {\em symmetric separable}
if these functions satisfy $f_1=\cdots=f_p$. It is called {\em completely separable}
if $f(x^1,\dots,x^p)=\sum_{k=1}^p\sum_{i=1}^d f_{k,i}(x^k_i)$ for some given
functions $f_{k,i}:\R\rightarrow\R$, and {\em symmetric completely separable}
if these functions satisfy $f_{1,1}=\cdots=f_{p,d}$.

The corresponding specializations of the type partition problem are similarly defined.\break
The functions are given by oracles that return the function values on queried arguments.

\vskip.2cm
In this article we study the computational complexity of these problems under
various assumptions on their natural parameters $p,d,a,t$. In particular, we discuss the
parameterized complexity of the problem, see \cite{CFKLMPPS}. Scaling up rational data if necessary,
we assume for simplicity that all data is integer, that is, the attribute matrix has integer
entries and all functions take on integer values on integer arguments. The running times of our
algorithms are in terms of the number of arithmetic operations and oracle queries they perform.

\vskip.2cm
First we note that with $d$ variable,
even with $p=2$ and $a=\|A\|_\infty=1$,
or $p$ variable, even with $d=1$ and
$a$ given in unary,
the problem requires exponential time, see Proposition \ref{exponential}.

Second we note that when $a$ is large, that is, the attribute matrix $A$ is binary encoded, even
with fixed $d=1$ and $p=2$, even the symmetric completely separable free problem with quadratic
convex functions of the form $f_{k,1}(x)=(x-b)^2$, is NP-hard, see Proposition \ref{NP_hard}.

In contrast, if the function $f:\R^{d\times p}\rightarrow\R$ is concave, then even the general
non-separable problem, even with $A$ binary encoded, is polynomial time solvable for every
fixed $p,d$. This follows from a geometric theory developed in \cite{HOR,Onn,OR,OSc},
briefly discussed in Section \ref{Discussion}.

\vskip.2cm
Here we establish the following. First, in Theorem \ref{general}, we show that for {\em any}
function $f:\R^{d\times p}\rightarrow\R$, even the general non-separable problem is solvable
in polynomial time for every fixed $p,d$ when $A$ is encoded in unary. Second, in
Theorem \ref{separable}, we show that for every fixed number $t$ of agent types,
the separable problem is solvable in polynomial time even if the number $p$ of parts and
attribute dimension $d$ are variable. Third, in Theorem \ref{completely_separable}, we show
that the completely separable bounded-shape problem with convex functions, is fixed-parameter
tractable on $p,d,a$ (see \cite{CFKLMPPS}), that is, solvable in time
which is a computable function of $p,d,a$ times a polynomial in $n$.
The proof of this result uses recent results on sparse integer programming
in variable dimension from \cite{EHKKLO,KLO,KO} which are the culmination of a theory initiated
in \cite{DHOW} and developed in \cite{Onn}, and whose origins could be traced back to \cite{Stu}.
We also establish analogous, usually stronger, results for the simpler type partition problem.
The next table outlines some of our results
(the precise statements are given in Section \ref{Statements}):

\vskip.2cm
\begin{tabular}{||c||c|c|c||}
  \hline
  \hline
& general & separable & convex completely separable \\
  \hline
  \hline
vector partition & $O\left(n^{(d+1)p+1}a^{dp}\right)$ & $O(dpn^{2t})$ & $(a+1)^{h(d,p)}\poly(n)$ \\
type partition &  $O\left(n^{(t+1)p+1}\right)$ & $O(pn^{2t})$ & $\poly(pt)\log(n)$ \\
  \hline
  \hline
\end{tabular}

\vskip.2cm
In Section \ref{Statements} we give the precise statements of the above theorems and their proofs.
In Section \ref{Application} we provide some motivation for the problem by briefly discussing a
recent application to testing for COVID-19. Finally, in Section \ref{Discussion},
we discuss the geometric method of solving the problem for concave objective functions,
and some of the many remaining open problems.

\section{Statements and proofs}
\label{Statements}

\subsection{The general problem: fixed number of parts and attribute dimension}

First we note that with $d$ variable,
even with $p=2$ and $a=\|A\|_\infty=1$,
or $p$ variable, even with $d=1$ and
$a=\|A\|_\infty$ given in unary,
the problem requires exponential time.
\bp{exponential}
The vector partition problem requires exponential time as follows: (1) with variable $d$, fixed $p=2$, and
$0-1$ matrix $A$, so with fixed
$a=\|A\|_\infty=1$,
the free problem, with $B_1=B_2=\{0,1,\dots,n\}$, needs time $\Omega(2^n)$;
(2) with variable $p$ and fixed $d=1$,
and $a$ given in unary, the single-shape problem,
with $B_k=\{1\}$ for $k=1,\dots,p$, needs time $\Omega(n!)$.
\ep
\bpr
For (1), let $d:=n$ and let $A^j:={\bf 1}_j$ be the $j$-th unit vector in $\R^d$
for $j=1,\dots,n$. (Note that the number of agent types is $t=n$ hence also variable.)
Then, for a $2$-partition $\pi$, the first column of the matrix
$A(\pi):=\left(\sum_{j\in\pi_1}A^j,\sum_{j\in\pi_2}A^j\right)\in\{0,1\}^{n\times 2}$ is
the indicating vector $A^1(\pi)=\sum_{j\in\pi_1}{\bf 1}_j$ of $\pi_1\subseteq[n]$. Thus, the
$2$-partitions $\pi$ of $[n]$ give $2^n$ distinct matrices $A(\pi)$ with first columns
all vectors in $\{0,1\}^n$. So the function $f$ can have arbitrary distinct values
$f\left(\sum_{j\in\pi_1}A^j,\sum_{j\in\pi_2}A^j\right)=f(A(\pi))$ on the matrices $A(\pi)$
of all distinct $2$-partitions.

For (2), let $p:=n$ and let $A:=[1,2,\dots,n]$. (Again, $t=n$ is variable.)
Then, the $n$-partitions with $|\pi_k|=1$ for all $k$ correspond to the $n!$ permutations
$\pi=(\pi_1,\dots,\pi_n)$ of $A$, so the function $f$ can have arbitrary distinct values
$f(\pi_1,\dots,\pi_n)$ on these distinct permutations.

So if an algorithm trying to solve the partition problem fails to query the oracle
of $f$ on one of the $2^n$ matrices $A(\pi)$ in the first case, or one of the $n!$ permutations
in the second case, it cannot tell whether the minimum value of $f$ is attained at the missing
matrix or the missing permutation or not, so it may fail to correctly solve the problem.
So the algorithm must query on all $2^n$ matrices $A(\pi)$ in the first case
and on all $n!$ permutations in the second case.
\epr

In contrast, when both $p,d$ are fixed, and $A$ is encoded in unary, we have the following.

\bt{general}
For the general partition problems with fixed number $p$ of parts we have:
\begin{enumerate}
\item
For any fixed $d$, the vector partition problem with matrix $A\in\Z^{d\times n}$ with
$a:=\|A\|_\infty$, $B_1,\dots,B_p\subseteq[n]$, and any function $f:\Z^{d\times p}\rightarrow\Z$,
is solvable in time $O\left(n^{(d+1)p+1}a^{dp}\right)$, that is, in time
polynomial in the number $n$ of agents and in the unary encoding of $a$.
\item
For any fixed $t$, the type partition problem with $t$ types and with an arbitrary function
$g:\Z^{t\times p}\rightarrow\Z$, can be solved in time $O\left(n^{(t+1)p+1}\right)$
polynomial in the number $n$ of agents.
\end{enumerate}
\et
\bpr
We begin with part 1. We augment the given attribute matrix $A$ with an additional
row indexed $0$, and we set $A_0^1:=\cdots:= A_0^n:=1$. We let $V\subset\Z^{(d+1)\times p}$
be the set of integer $(d+1)\times p$ matrices $v$ which satisfy $0\leq v_{0,k}\leq n$
and $0\leq |v_{i,k}|\leq na$ for all $i\in[d]$ and $k\in[p]$.
We denote by ${\bar A}\in\Z^{d\times n}$ and ${\bar v}\in\Z^{d\times p}$ the
submatrices of $A$ and $v$, respectively, obtained by deleting their $0$-th row.
In particular, $\bar A$ is the original attribute matrix. We now reduce the vector partition
problem to that of finding a shortest directed path in an acyclic directed graph $D$ where
each edge has a length. We construct $D$ as follows. There is one special vertex $s$, and
all other vertices are labeled by pairs $[j,v]$ with $j\in\{0,1,\dots,n\}$ and $v\in V$.
For $j=1,\dots,n$ and $v,w\in V$ there is an edge, of length $0$, from $[j-1,v]$ to $[j,w]$,
if and only if $w=v+A^j\otimes{\bf 1}_k$ for some $k\in[p]$, with ${\bf 1}_k$ the $k$-th unit
vector in $\R^p$, that is, $w$ is obtained by adding $A^j$ to the $k$-th column of $v$.
For each $v\in V$ with $v_{0,k}\in B_k$ for all $k\in[p]$ there is an edge from $[n,v]$ to $s$ of
(possibly negative) length $f({\bar v})$. Let ${\bf 0}\in\Z^{(d+1)\times p}$ be the $0$ matrix.

For any $p$-partition $\pi$ with $|\pi_k|\in B_k$ for all $k\in[p]$,
we attach a path from $[0,{\bf 0}]$ to $s$ in $D$,
\begin{equation}\label{path1}
[0,v^0={\bf 0}]\rightarrow[1,v^1]
\rightarrow\cdots\rightarrow[n,v^n]\rightarrow s\ ,
\end{equation}
where, for $j=1,\dots,n$, letting $k(j)$ be such that $j\in\pi_{k(j)}$, we set
$v^j:=v^{j-1}+A^j\otimes{\bf 1}_{k(j)}$. Note that each entry $v^j_{i,k}$ of each $v^j$ is
obtained by summing at most $n$ entries of $A$ and hence has nonnegative value bounded by $n$ for
$i=0$ and absolute value bounded by $n\|A\|_\infty=na$ for $i=1,\dots,d$. So $v^j\in V$ for all $j$.
Moreover, for $k=1,\dots,p$, the $k$-th column of $v^n$ is equal to $\sum_{j\in\pi_k}A^j$.
In particular, $v^n_{0,k}=\sum_{j\in\pi_k}A^j_0=|\pi_k|\in B_k$
and hence there is an edge from $[n,v^n]$ to $s$. So this is indeed a path in $D$.
Furthermore, the length of the path, which is the length $f({\bar v^n})$ of the edge
from $[n,v^n]$ to $s$, is $f(\sum_{j\in\pi_1}{\bar A}^j,\dots,\sum_{j\in\pi_p}{\bar A}^j)$,
which is equal to the objective value of the partition $\pi$ in the vector partition problem.

Conversely, from each path as in \eqref{path1} we read off a partition
$\pi=(\pi_1,\dots,\pi_p)$ as follows. For $j=1,\dots,n$ there is a $k(j)\in[p]$ such that
$v^j:=v^{j-1}+A^j\otimes{\bf 1}_{k(j)}$. We let $\pi_k:=\{j:k(j)=k\}$. Since there is an edge from
$[n,v^n]$ to $s$ we have $|\pi_k|\in B_k$ for $k\in[p]$. Moreover, the objective value of $\pi$
in the vector partition problem is equal to the length $f({\bar v}^n)$ of the path.

So finding a minimum cost vector partition reduces to finding a shortest directed path
in $D$ from $[0,{\bf 0}]$ to $s$ and reading off an optimal partition as explained above.

Now, a shortest path in an acyclic directed graph can be found in time
which is linear in the number of edges of $D$ (see e.g. \cite{Sch}),
which, as claimed, is $O(np|V|)=O\left(n^{(d+1)p+1}a^{dp}\right)$.

For part 2, as explained in the introduction, the type partition problem with $t$ types can be
encoded as a vector partition problem with attribute dimension $d=t$ and $0-1$ attribute
matrix $A$, so with $a=1$, implying part 2 about the type partition problem as well.
\epr

\subsection{The separable problem: fixed number of agent types}

Next we consider fixed number $t$ of agent types in both vector and type partition problems.

\bt{separable}
The following statements hold even with a variable number $p$ of parts:
\begin{enumerate}
\item
For any fixed $t$, the separable type problem with $t$ types is solvable in time $O(pn^{2t})$;
\item
For any fixed $t$, the separable vector problem with $t$ types is solvable in time $O(dpn^{2t})$;
\item
For any fixed $d$ and $a$, the separable vector partition problem with an
integer $d\times p$ attribute matrix $A$ satisfying $\|A\|_\infty\leq a$
is solvable in polynomial time $O(pn^{2(2a+1)^d})$.
\end{enumerate}
\et
\bpr
We begin with part 1 about the type partition problem. Let $\tau=(\tau_1,\dots,\tau_t)$
be the given partition into types and let $n_i:=|\tau_i|$. Let $g_k:\R^t\rightarrow\R$ be the given
functions for $k=1,\dots,p$ so that the cost of a $p$-partition $\pi=(\pi_1,\dots,\pi_p)$
is $\sum_{k=1}^p g_k\left(|\pi_k\cap\tau|\right)$. Define
$$V\ :=\ \left\{(v_1,\dots,v_t)\ :\ v_i\in\{0,1,\dots,n_i\},\ i=1,\dots,t\right\}
\ \subseteq\ \{0,1,\dots,n\}^t\ \subset\ \R^t\ .$$
We again reduce the problem to that of finding a shortest directed path in an acyclic
directed graph $D$. We construct this graph as follows. The vertices are
labeled by pairs $[k,v]$ with $k\in\{0,1,\dots,p\}$ and $v\in V$. There is an edge from
$[k-1,v]$ to $[k,w]$ if and only if for $i=1,\dots,t$ we have $w_i-v_i\geq 0$, and
$\sum_{i=1}^t(w_i-v_i)\in B_k$, where $B_1,\dots,B_p$ are the input subsets in the
partition problem. The (possibly negative) length of this edge is $g_k(w-v)$.
Consider any $p$-partition $\pi$ with $|\pi_k|\in B_k$ for $k\in[p]$.
For $k=1,\dots,p$ let $x^k:=|\pi_k\cap\tau|$.
This partition is encoded as the following directed path in $D$
from $[0,(0,\dots,0)]$ to $[p,(n_1,\dots,n_t)]$,
$$\hskip-0.2cm[0,v^0=(0,\dots,0)]\rightarrow[1,v^1=x^1]\rightarrow[2,v^2=x^1+x^2]
\rightarrow\cdots\rightarrow[p,v^p=x^1+\cdots+x^p=(n_1,\dots,n_t)]\ .$$
All edges in the path exist, since for all $k\in[p]$ we have
$\sum_{i=1}^t(v^k_i-v^{k-1}_i)\in B_k$, because
$$\sum_{i=1}^t(v^k_i-v^{k-1}_i)=\sum_{i=1}^t x^k_i=\sum_{i=1}^t|\pi_k\cap\tau_i|=|\pi_k|\ .$$
Further, the length of this path is the following expression, which is precisely the cost of $\pi$,
$$\sum_{k=1}^p g_k(v^k-v^{k-1})\ =\ \sum_{k=1}^p g_k(x^k)
\ =\ \sum_{k=1}^p g_k\left(|\pi_k\cap\tau|\right)\ .$$
Conversely, consider any directed path in $D$ from $[0,(0,\dots,0)]$ to $[p,(n_1,\dots,n_t)]$,
$$\hskip-0.2cm[0,v^0=(0,\dots,0)]\rightarrow[1,v^1]\rightarrow[2,v^2]
\rightarrow\cdots\rightarrow[p,v^p=(n_1,\dots,n_t)]\ .$$
For $k=1,\dots,p$ let $x^k:=v^k-v^{k-1}$. Then $x^k\geq 0$ for all $k$ and
$x^1+\cdots+x^p=v^p=(n_1,\dots,n_t)$, so for $i=1,\dots,t$ we have $x^1_i+\cdots+x^p_i=n_i$.
Letting $\pi=(\pi_1,\dots,\pi_p)$ be a $p$-partition of $[n]$ where for $k=1,\dots,p$ we let
$\pi_k$ consist of $x^k_i$ agents of type $i$ for each $i=1,\dots,t$, we have that
$|\pi_k\cap\tau|=x^k$, $|\pi_k|=\sum_{i=1}^t x^k_i\in B_k$,
and the cost of $\pi$ is precisely the path length,
$$\sum_{k=1}^p g_k\left(|\pi_k\cap\tau|\right)
\ =\ \sum_{k=1}^p g_k(x^k)\ =\ \sum_{k=1}^p g_k(v^k-v^{k-1})\ .$$

So, finding a minimum cost type partition reduces to finding a shortest directed path in $D$ from
$[0,(0,\dots,0)]$ to $[p,(n_1,\dots,n_t)]$ and reading off an optimal partition as explained above.

Now, a shortest path in an acyclic directed graph can be found in time linear in the number of
edges of $D$ (see again e.g. \cite{Sch}), which is $O(p|V|^2)=O(pn^{2t})$, proving part 1.

We proceed with part 2 about the vector partition problem with $t$ types. We now show that it
can be encoded as a suitable type partition problem. Let $A\in\Z^{d\times n}$ be the attribute
matrix, $B_1,\dots,B_p\subseteq[n]$, and $f_k:\R^d\rightarrow\R$
for $k=1,\dots,p$ be the functions, forming the data for the vector partition problem.
Relabeling agents if necessary, assume $A^1,\dots,A^t$ are the $t$ distinct attribute vectors,
and for $i=1,\dots,t$ let $\tau_i\subseteq[n]$ be the set of agents of type $i$, that is,
with $A^j=A^i$ for every $j\in\tau_i$. Also let $n_i:=|\tau_i|$. For $k=1,\dots,p$ define
a function $g_k:V\rightarrow\R$ by $g_k(v_1,\dots,v_t):=f_k\left(\sum_{i=1}^t v_iA^i\right)$.
Then the cost of a $p$-partition $\pi$ satisfies
$$\sum_{k=1}^p f_k\left(\sum_{j\in\pi_k}A^j\right)
\ =\ \sum_{k=1}^p f_k\left(\sum_{i=1}^t\sum_{j\in\pi_k\cap\tau_i}A^j\right)
\ =\ \sum_{k=1}^p f_k\left(\sum_{i=1}^t|\pi_k\cap\tau_i|A^i\right)
\ =\ \sum_{k=1}^p g_k\left(|\pi_k\cap\tau|\right)\ .$$
This encodes the vector problem as a type problem. Now, evaluating any $g_k(v_1,\dots,v_t)$
takes $(2t-1)d=O(d)$ arithmetic operations to compute $\sum_{i=1}^t v_iA^i$ plus one
call to the oracle of $f_k$, and we need such an evaluation to compute the length
of each edge of the graph $D$. Using part 1, the total time for solving
the vector partition problem is $O(dpn^{2t})$, proving part 2.

For part 3, note that if $d$ and $a$ are fixed then the number of possible distinct columns of
$A$ satisfies $t\leq (2a+1)^d$, so part 3 follows from part 2 on fixed number $t$ of types.
\epr

\subsection{The completely separable problem: fixed-parameter tractability}

We begin by noting that when $a$ is large, that is, the attribute matrix $A$ is binary encoded,
even with fixed $d=1$ and fixed $p=2$, even the symmetric completely separable free
partition problem with quadratic convex functions of the simple form $f_{k,1}(x)=(x-b)^2$, is NP-hard.

\bp{NP_hard}
The symmetric completely separable problem is NP-hard even with $d=1$, $p=2$,
convex functions $f_{k,1}(x)=(x-b)^2$, and the free case, with $B_1=B_2=\{0,1,\dots,n\}$.
\ep
\bpr
It is well known to be NP-complete to decide if given numbers $a_1,\dots,a_n\in\Z_+$
can be partitioned to two parts of equal sums. Let $A:=[a_1,\dots,a_n]$, let
$b:={1\over2}\sum_{j=1}^na_j$, and let the functions be $f_{1,1}(x):=f_{2,1}(x):=(x-b)^2$.
Then the cost of $2$-partition $\pi=(\pi_1,\pi_2)$ is
$(\sum_{j\in\pi_1}a_j-b)^2+(\sum_{j\in\pi_2}a_j-b)^2$, and is nonnegative and equals to $0$
if and only if $\sum_{j\in\pi_1}a_j=\sum_{j\in\pi_2}a_j$. Thus, the optimal value of the
symmetric completely separable free partition problem is $0$ if and only if
$a_1,\dots,a_n$ can be partitioned to two parts of equal sums.
\epr

\vskip.2cm
In what follows we make use of nonlinear integer programming in standard form,
\begin{equation}\label{IP}
\min\{f(x)\ :\ Ax=b,\ \ l\leq x\leq u,\ \ x\in\Z^n\}\ ,
\end{equation}
with $A\in\Z^{m\times n}$, $b\in\Z^m$, $l,u\in\Z^n$, and $f:\R^n\rightarrow\R$ a separable convex
function, that is, given by  $f(x)=\sum_{j=1}^n f_j(x_j)$ with $f_j:\R\rightarrow\R$ a\
univariate convex function for all $j$. As mentioned in the introduction,
we assume that all $f_j$ take on integer values on integer arguments.
We also denote by $L:=\log(\|u-l\|_\infty+1)$ the bit complexity of the lower and upper bounds.

Unfortunately, as is well known, the problem is generally NP-hard even for linear objective
functions. Moreover, the classical result of Lenstra \cite{Len} that the problem with linear
objectives, parameterized by the number of variables, is fixed-parameter tractable, cannot help
us, since our integer programs below involve a variable number of variables. So we need to
review and use one early and one recent result on integer programming in variable dimension.

The first result we need, which is quite well known by now, from \cite{HS}, concerns the case
when the matrix $A$ is totally unimodular, that is every subdeterminant of $A$ is $-1,0,1$.

\bp{TU}{\rm \cite{HS}}
Program \eqref{IP} with $A$ totally unimodular is solvable in time $\poly(n)L$.
\ep

Next, we need to discuss recent results of \cite{EHKKLO,KLO,KO} on fixed-parameter tractability
of sparse integer programming. For this we need some more terminology.
The tree-depth of a graph $G=(V,E)$ is defined as follows. The {\em height} of a rooted
tree is the maximum number of vertices on a path from the root to a leaf. A rooted tree
on $V$ is {\em valid} for $G$ if for each edge $\{i,j\}\in E$ one of $i,j$ lies on the
path from the root to the other. The {\em tree-depth} $td(G)$ of $G$ is the smallest height
of a rooted tree which is valid for $G$, see also \cite{NO}. For instance, if $G=([2m],E)$ is a
perfect matching with $E=\{\{i,m+i\}:i\in[m]\}$ then its tree-depth is $3$ where a
tree validating it rooted at $1$ has edge set $E\uplus\{\{1,i\}:i=2,\dots,m\}$.
Next, the graph of an $m\times n$ matrix $A$ is the graph $G(A)$ on $[n]$ where ${j,k}$ is
an edge if and only if there is an $i\in[m]$ such that $A_i^jA_i^k\neq 0$. The {\em tree-depth}
of $A$ is the tree-depth $td(A):=td(G(A))$ of its graph. The following result asserts that the
integer programming problem \eqref{IP} (with separable convex and in particular linear objective)
is solvable in fixed-parameter tractable time (see \cite{CFKLMPPS}) parameterized by the
{\em numeric measure} $a:=\|A\|_\infty$ and {\em sparsity measure} $d:=td(A^T)$ of the
transpose $A^T$ of the matrix defining the program. Moreover, for any fixed $d$ it is solvable
in polynomial time even when $a$ is a variable part of the input which is encoded in unary.
In the proposition and theorem below, $g$ and $h$ are some computable functions of the parameters.
\bp{TD}{\rm \cite{KO}}
Program \eqref{IP} is solvable in fixed-parameter tractable time
$$(a+1)^{g(d)}\poly(n)L\ .$$
\ep

We are now is position to show that the completely separable bounded-shape vector partition
problem with convex functions, parameterized by $p,d,a$, is fixed parameter-tractable.
The subsets are intervals $B_k:=\{z\in\Z:l_k\leq z\leq u_k\}$ defined by
given bounds $l,u\in\Z_+^p$. We denote $|\pi|:=(|\pi_1|,\dots,|\pi_p|)$,
so the admissible partitions are those satisfying $l\leq|\pi|\leq u$.

\bt{completely_separable}
For the completely separable partition problems with convex functions we have:
\begin{enumerate}
\item
The completely separable vector problem, with matrix $A\in\Z^{d\times n}$
with $a:=\|A\|_\infty$, bounds $l,u\in\Z_+^p$, and convex functions $f_{k,i}:\Z\rightarrow\Z$,
is solvable in time $(a+1)^{h(d,p)}\poly(n)$. Hence it is fixed-parameter
tractable on $p,d,a$, and for fixed $p,d$ it is solvable in time polynomial in the
unary encoding of $a$ and polynomial in $n$ of degree independent of $p,d$.

\item
The completely separable type problem, with convex functions $g_{k,i}:\Z\rightarrow\Z$, is solvable
in time $\poly(pt)\log(n)$, even with variable the number $p$ of parts and number $t$ of types.
\end{enumerate}
\et
\bpr
We begin with part 1 about the vector partition problem. We augment again the given attribute
matrix $A$ with an additional row indexed $0$, and we set $A_0^1:=\cdots:= A_0^n:=1$. We now
construct an integer program as follows. There are binary variables $x_{k,j}$ for $k=1,\dots,p$
and $j=1,\dots,n$ with the interpretation that $x_{k,j}=1$ indicates that agent $j$ goes to
part $\pi_k$ of the sought partition. There are integer variables $y_{k,i}$
for $k=1,\dots,p$ and $i=0,\dots,d$ with the interpretation that in the partition $\pi$
determined by the $x_{k,j}$, the attribute sums satisfy $y_{k,i}=\sum_{j\in\pi_k}A^j_i$.
Here is the program, followed by further explanation:

$$\min\sum_{k=1}^p\sum_{i=1}^df_{k,i}(y_{k,i})$$
\begin{equation}\label{IP1}
\sum_{j=1}^n A^j_i x_{k,j} - y_{k,i}\ =\ 0\ ,\quad\quad k=1,\dots,p\ ,\quad\quad i=0,\dots,d\ ,
\end{equation}
\begin{equation}\label{IP2}
\sum_{k=1}^p x_{k,j}\ =\ 1\ ,\quad\quad j=1,\dots,n\ ,
\end{equation}
$$0\leq x_{k,j}\leq 1,\ \ l_k\leq y_{k,0}\leq u_k,\ \ -na\leq y_{k,i}\leq na,\ \
x_{k,j},y_{k,0},y_{k,i}\in\Z,\ \ k\in[p],\ \ i\in[d],\ \ j\in[n]\ .$$

\vskip.2cm
On the one hand, consider any feasible solution $(x,y)$ of this program and define a tuple
$\pi=(\pi_1,\dots,\pi_p)$ by $\pi_k:=\{j:x_{k,j}=1\}$ for all $k$. Equations \eqref{IP2}
guarantee that $\pi$ is indeed a $p$-partition of $[n]$ and equations \eqref{IP1}
guarantee that indeed $y_{k,i}=\sum_{j\in\pi_k}A^j_i$ for all $k,i$. In particular,
$|\pi_k|=\sum_{j\in\pi_k}A^j_0=y_{k,0}$ for all $k$, so $l\leq|\pi|\leq u$ and $\pi$
is an admissible partition. Moreover the objective function value of this solution
is equal to $\sum_{k=1}^p\sum_{i=1}^df_{k,i}(\sum_{j\in\pi_k}A^j_i)$ and is indeed
the objective function value of the partition $\pi$ in the vector partition problem.

On the other hand, consider any $p$-partition $\pi=(\pi_1,\dots,\pi_p)$ of $[n]$ with
$l\leq|\pi|\leq u$. Define $(x,y)$ as follows. For all $k,j$ define $x_{k,j}:=1$ if $j\in\pi_k$
and $x_{k,j}:=0$ otherwise. Clearly $x$ satisfies equations \eqref{IP2}. Now define $y_{k,i}$
for all $k,i$ by equations \eqref{IP1}, which then automatically hold. Further, for all $k$
we have that $|y_{k,0}|=\sum_{j\in\pi_k}A^j_0=|\pi_k|$ so $l_k\leq y_{k,0}\leq u_k$, and
for $i=1,\dots,d$ we have that $|y_{k,i}|\leq\sum_{j\in\pi_k}|A^j_i|\leq na$, so all $y_{k,i}$
satisfy the bounds. Therefore $(x,y)$ is a feasible solution of the integer program.
Moreover, the objective function value of the partition $\pi$ in the vector partition problem
is equal to $\sum_{k=1}^p\sum_{i=1}^df_{k,i}(y_{k,i})$ and is indeed the
objective function value of the solution $(x,y)$ in the integer program.
This shows that the vector partition problem reduces to the above integer program.

Now consider the matrix $B$ expressing equations \eqref{IP1},\eqref{IP2} and its transpose $B^T$.
The columns of $B$ are indexed by the variables. Let us index the equations and the rows of
$B$ by $r_{1,0},\dots,r_{p,d}$ corresponding to equations \eqref{IP1} and $s_1,\dots,s_n$
corresponding to equations \eqref{IP2}. Let $T$ be the tree on the $r_{k,i}$ and $s_j$ which
is rooted at $r_{1,0}$ and consists of the path $(r_{1,0},\dots,r_{p,d})$ together with the
leaves $s_1,\dots,s_n$, each connected to $r_{p,d}$. The height of $T$ is Clearly $(d+1)p+1$.
We now show that $T$ is valid for $G(B^T)$. For this we need to show that if two equations share
a variable then one lies on the path in $T$ from its root to the other. Since all the $r_{k,i}$
lie on a common path, it is clear that this condition holds for any $r_{k,i},r_{{\bar k},{\bar i}}$.
It is also clear that this condition holds for any $s_j,s_{\bar j}$ since any two such distinct
equations involve disjoint sets of variables. Finally, the condition also holds
for any two equations $r_{k,i}$, $s_j$ since $r_{k,i}$ lies on the path from the root
$r_{1,0}$ to $s_j$. So the tree $T$ is indeed valid for $G(B^T)$.

Therefore the tree-depth of $B^T$ satisfies $td(B^T)\leq (d+1)p+1$. Also, it is clear that
the matrix $B$ satisfies $\|B\|_\infty=\|A\|_\infty=a$. Furthermore, all variables are
bounded below by $-na$ and above by $na$ so $L=\log(\|u-l\|_\infty+1)=\log(2na+1)$.
Finally, the number of variables is $p(n+d+1)$. Substituting these in Proposition \ref{TD},
we find that, for some suitable computable function $h$, the running time is
bounded by $(a+1)^{h(d,p)}\poly(n)$, proving part 1.

\vskip.2cm
We proceed with part 2 about the type partition problem. As explained in the introduction,
the problem with $t$ types can be encoded as a vector partition problem with attribute dimension
$d=t$ and $0-1$ attribute matrix $A$, so with $a=1$. Plugging this into the time bound for the
vector partition problem gives $2^{h(t,p)}\poly(n)$. However, for the type partition problem,
which is easier, we can do much better. Let then $\tau=(\tau_1,\dots,\tau_t)$ be the
given type partition with $n_i:=|\tau_i|$ for $i=1,\dots,t$. We construct again
a suitable integer program, reminiscent of but different than the one above. In particular,
the number of variables is independent of $n$, and the variables are not binary.
For $k=1,\dots,p$ and $i=1,\dots,t$ we now have variables $x_{k,i}$ and variables $y_{k}$
with the interpretation that $x_{k,i}=|\pi_k\cap\tau_i|$ and $y_k=|\pi_k|$ in the sought
partition $\pi$. Here is the program, followed by further explanation:
$$\min\sum_{k=1}^p\sum_{i=1}^t g_{k,i}(x_{k,i})$$
\begin{equation}\label{IP3}
\sum_{i=1}^t x_{k,i}-y_k\ =\ 0\ ,\quad\quad k=1,\dots,p\ ,
\end{equation}
\begin{equation}\label{IP4}
\sum_{k=1}^p x_{k,i}\ =\ n_i\ ,\quad\quad i=1,\dots,t\ ,
\end{equation}
$$0\leq x_{k,i}\leq\min\{u_k,n_i\}\ ,\quad l_k\leq y_k\leq u_k\ ,\quad
x_{k,i},y_k\in\Z\ ,\quad k\in[p]\ ,\quad i\in[d]\ .$$
On the one hand, consider any feasible solution $(x,y)$. Then for all $i$, equations
\eqref{IP4} assert that $\sum_{k=1}^p x_{k,i}=n_i=|\tau_i|$, so we can define a $p$-partition
$\pi=(\pi_1,\dots,\pi_p)$ where for all $k$ we let $\pi_k$ consist of $x_{k,i}$ agents
of type $i$, so that indeed $|\pi_k\cap\tau_i|=x_{k,i}$. Then for all $k$, equations
\eqref{IP3} imply $y_k=\sum_{i=1}^t|\pi_k\cap\tau_i|=|\pi_k|$, so the bounds on $y_k$ imply
$l_k\leq|\pi_k|\leq u_k$ and $\pi$ is an admissible partition. Moreover, the objective function
value of this solution is equal to $\sum_{k=1}^p\sum_{i=1}^t g_{k,i}(|\pi_k\cap\tau_i|)$
and is indeed the value of $\pi$ in the type partition problem.

On the other hand, consider any $p$-partition $\pi=(\pi_1,\dots,\pi_p)$ satisfying
$l\leq|\pi|\leq u$. Define $(x,y)$ as follows. For all $k,i$ let
$x_{k,i}:=|\pi_k\cap\tau_i|\leq\min\{u_k,n_i\}$ which implies that
$\sum_{k=1}^p x_{k,i}=\sum_{k=1}^p|\pi_k\cap\tau_i|=|\tau_i|=n_i$ so
equations \eqref{IP4} hold, and define $y_k$ by equations \eqref{IP3} which then hold
and give $y_k=\sum_{i=1}^t|\pi_k\cap\tau_i|=|\pi_k|$, so the bounds on $|\pi_k|$ imply
$l_k\leq y_k\leq u_k$. Therefore $(x,y)$ is a feasible solution of the integer program.
Moreover, the objective function value of the partition $\pi$ in the type partition problem is
equal to $\sum_{k=1}^p\sum_{i=1}^t g_{k,i}(x_{k,i})$ and is indeed the objective function
value of the solution $(x,y)$ in the integer program.
This shows that the type partition problem reduces to the above integer program.

Now consider the matrix $C$ expressing equations \eqref{IP3},\eqref{IP4}.
We use the following well known sufficient condition for total unimodularity (see e.g. \cite{Sch}).
Suppose a matrix has $-1,0,1$ entries, at most two nonzero entries per column, and the rows can be
$2$-partitioned so that, for every column with two nonzero entries, if these entries have the same
sign then they lie in rows in opposite parts, whereas if they have opposite signs then they lie in
rows in the same part. Then the matrix is totally unimodular. Now, all entries of $C$ are $-1,0,1$.
Each variable $y_k$ appears in one equation so the corresponding column has one nonzero entry.
Each variable $x_{k,i}$ appears in two equations, one from \eqref{IP3} and one from \eqref{IP4},
so the corresponding column has exactly two nonzero entries of the same sign, one from the
set of rows corresponding to equations \eqref{IP3} and one from the set of rows corresponding
to equations \eqref{IP4}, which form the desired $2$-partition of rows.
So by the above sufficient condition, $C$ is totally unimodular.

Also, all variables are bounded below by $0$ and above by $n$ so
$L=\log(\|u-l\|_\infty+1)\leq\log(n+1)$. Finally, the number of variables is $p(t+1)$.
Substituting these in Proposition \ref{TU}, the running time
is found to be bounded by $\poly(pt)\log(n)$, proving part 2.
\epr

Two remarks are in order here. First, note that for fixed $p,d$, the running time for the vector
partition problem in both Theorem \ref{general} part 1 and Theorem \ref{completely_separable}
part 1 is polynomial in $n$ and $a$. The former is stronger in that it holds for the general,
non-separable non-convex problem, and the degree of the polynomial in $a$ is $dp$ which is much
smaller than the degree $h(d,p)$ in the latter. But the latter is stronger in that the degree
of the polynomial in $n$ is independent of $p,d$ whereas it is $(d+1)p+1$ in the former.
Second, it is tempting, in the proof of Theorem \ref{completely_separable}, to try and encode the
completely separable vector partition problem as a completely separable type partition problem,
as done for the (non-completely) separable problems in the proof of Theorem \ref{separable}
part 2, to aim at a running time which is polynomial also in $p,d$ and independent of $a$.
However, this does not work, as can be expected from the NP-hardness of the completely
separable vector partition problem asserted in Proposition \ref{NP_hard}.

\section{Application}
\label{Application}

Here we very briefly discuss the interesting application to testing for COVID-19 from \cite{EBBA}.
Patients (agents) exposed to infected individuals are to be tested for the virus in a facility
with limited testing capacity. They are to be partitioned into groups: the first group
consists of patients not to be tested at all; the second group consists of patients to be each
tested individually; the other groups consist of patients to be tested according to the
so-called Dorfman testing scheme: all patients in each such group are tested together
(by mixing their specimens); if the result is negative then they are all declared negative,
whereas if it is positive then each is tested again individually. The patient population is
assumed heterogeneous: with each patient are associated two parameters: the {\em positivity risk}
which depends on the nature of exposure to an infected individual, such as the exposing activity,
its duration, proximity, and environment; and the gap between {\em pre-intervention harm} and
{\em post-intervention harm} which estimate, respectively, the potential to spread the infection
if undetected or detected and treated (such as by isolation, quarantine, or symptom monitoring),
and which depend on the professional network (such as students, teachers, health care professionals,
grocery store workers) which affects the potential to spread the infection. The authors indicate
that it suffices to approximate these parameters by three levels (low, average, high).
Thus, the agents come in nine types. Using these parameters and the tests true-positive and
true-negative probabilities, they are able to formulate the problem of grouping patients for
testing so as to minimize expected harm, with expected number of tests not
exceeding capacity, as a suitable type partition problem, with $t=9$ agent
types and a variable number $p$ of parts (as in our Theorem \ref{separable}),
and describe a heuristic for solving it. See \cite{EBBA} for the precise details.

\section{Discussion}
\label{Discussion}

First, we very briefly discuss the geometric theory developed in \cite[Chapter 2]{Onn} and
\cite{HOR,OR,OSc}, which in particular enables to solve the vector partition problem with large,
binary encoded attributes, and concave functions, for every fixed $p,d$. This theory considers
minimization over sets $S\subset\Z^N$ where the objective is composite of the form
$f(Wx)$ where $W\in\R^{D\times N}$ and $f:\R^D\rightarrow\R$ is concave. It is assumed that
$f$ is given by an oracle, and that we are given an oracle for linear optimization over $S$
and a set $E\subset\R^N$ of edge-directions of the convex hull $\conv(S)$ of $S$
in $\R^N$. The framework involves the construction of the zonotope $Z\subset\R^D$ generated
by the projection $WE$ of $E$ by $W$ into $\R^D$, and the enumeration of its vertices,
which can be done in polynomial time for fixed $D$. Then a vertex $y$ of $Z$
minimizing $f$ is obtained, and finally, a point $x\in S$ projecting down to $y=Wx$
is determined. To specialize this framework to the vector partition problem, we let $N=pn$
and let $S\subset\{0,1\}^{p\times n}$ be the set of $0-1$ matrices with each column sum $1$,
which correspond to $p$-partitions of $[n]$. Linear optimization over $S$ is equivalent
to solving a transportation problem which is easy to do, and a set of edge-directions of
polynomial size can also be determined. We let $D=pd$ and then the attribute matrix $A$ for
the partition problem induces the matrix $W$ for the above framework. See \cite[Chapter 2]{Onn}
for the precise details of this framework, for the proof that these problems can be solved
in polynomial time even in the binary encoding of $W$, and for the description of the
specialization of this framework to the vector partition problem.

\vskip.2cm
Second, we discuss some of the many remaining open problems here. First, the results
of \cite{HOR} solve the problem with general concave function $f$ and binary encoded
attribute matrix $A$ for every fixed $p,d$ in strongly polynomial time $O(n^{dp^2})$.
It would be very interesting to know if the problem with a concave function is
in fact fixed-parameter tractable or W[1]-hard (see \cite{CFKLMPPS}) when $d$ is fixed and $p$
is the parameter or when $p$ is fixed and $d$ is the parameter.

Similarly, Theorem \ref{general} solves the problem with an arbitrary function $f$ and unary encoded
attribute matrix $A$ for every fixed $p,d$ in polynomial time $O(n^{(d+1)p+1}a^{dp})$. It would
again be very interesting to know if the problem is in fact fixed-parameter tractable or W[1]-hard
when $d$ is fixed and $p$ is the parameter or when $p$ is fixed and $d$ is the parameter.

Another interesting question, related to Theorem \ref{separable}, is if the separable type partition
problem with variable $p$, or the separable vector partition problem with variable $d$ and $p$,
are fixed-parameter tractable or W[1]-hard when parameterized by the number $t$ of types.

Last, it would be interesting to know if Theorem \ref{completely_separable} can be extended to show
that the (non-completely) separable problem is fixed-parameter tractable parameterized by $p,d,a$.

\section*{Acknowledgments}
S. Onn was supported by a grant from the Israel Science Foundation and the Dresner chair.


\begin{thebibliography}{}

\bibitem{BOT}
Babson, E.K., Onn, S., Thomas, R.R.:
The Hilbert zonotope and a polynomial time algorithm for universal Grobner bases.
Advances in Applied Mathematics 30:529--544 (2003)

\bibitem{COR}
Chakravarty, A.K., Orlin, J.B., Rothblum, U.G.:
Consecutive optimizers for a partitioning problem with applications to optimal
inventory groupings for joint replenishment.
Operations Research 33:820--834 (1985)

\bibitem{CFKLMPPS}
Cygan, M., Fomin, F.V., Kowalik, {\L}., Lokshtanov, D.,
Marx, D., Pilipczuk, M., Pilipczuk, M., Saurabh, S.:
Parameterized Algorithms. Springer (2015)

\bibitem{DHOW}
De Loera, J.A., Hemmecke, R., Onn, S., Weismantel, R.:
N-fold integer programming.
Discrete Optimization 5:231--241 (2008)

\bibitem{EHKKLO}
Eisenbrand, F., Hunkenschr\"oder, C., Klein, K.M., Kouteck\'y, M., Levin, A., Onn, S.:
An algorithmic theory of integer programming.
ArXiv:1904.01361 1--63 (2019)

\bibitem{EBBA}
El Hajj, H., Bish, D.R., Bish, E.K., Aprahamian, H.:
Screening multi-dimensional heterogeneous populations for infectious diseases
under scarce testing resources, with application to COVID-19.
Naval Research Logistics 1--18 (2021)
{\tt https://doi.org/10.1002/nav.21985}

\bibitem{HS}
Hochbaum, D.S., Shanthikumar, J.G.:
Convex separable optimization is not much harder than linear optimization.
Journal of the ACM 37:843--862 (1990)

\bibitem{HOR}
Hwang, F.K., Onn, S., Rothblum, U.G.:
A polynomial time algorithm for shaped partition problems.
SIAM Journal on Optimization 10:70--81 (1999)

\bibitem{KLO}
Kouteck\'y, M., Levin, A., Onn, S.:
A parameterized strongly polynomial algorithm for block structured integer programs.
Proceedings of ICALP 2018, Leibniz International Proceedings in Informatics, 107-85:1--14 (2018)

\bibitem{KO}
Kouteck\'y, M., Onn, S.:
Sparse Integer Programming is FPT.
Bulletin of the European Association for Theoretical Computer Science 134:69--71 (2021)

\bibitem{Len}
Lenstra, H.W., Jr.:
Integer programming with a fixed number of variables.
Mathematics of Operations Research 8:538--548 (1983)

\bibitem{NO}
Ne\v{s}et\v{r}il, J., Ossona de Mendez, P.:
Sparsity: Graphs, Structures, and Algorithms.
Algorithms and Combinatorics, Springer (2012)

\bibitem{Onn}
Onn, S.:
Nonlinear Discrete Optimization.
Zurich Lectures in Advanced Mathematics, European Mathematical Society (2010).
Available online at:\\
{\tt http://ie.technion.ac.il/$\sim$onn/Book/NDO.pdf}

\bibitem{OR}
Onn, S., Rothblum, U.G.:
Convex combinatorial optimization.
Discrete and Computational Geometry 32:549--566 (2004)

\bibitem{OSc}
Onn, S., Schulman, L.:
The vector partition problem for convex objectice functions.
Mathematics of Operations Research 26:583--590 (2001)

\bibitem{OSt}
Onn, S., Sturmfels, B.:
Cutting corners.
Advances in Applied Mathematics 23:29--48 (1999)

\bibitem{Sch}
Schrijver A.:
Combinatorial Optimization. Springer (2003)

\bibitem{Stu}
Sturmfels, B.:
Gr\"obner Bases and Convex Polytopes.
University Lecture Series, American Mathematical Society (1996)

\end{thebibliography}
\end{document}